\documentclass[english,aps,prl,reprint,nofootinbib,nobibnotes,notitlepage,preprintnumbers,showpacs]{revtex4-1}
\usepackage{amsmath, amssymb, amsthm, braket}
\usepackage{graphicx}
\usepackage{subfigure}
\usepackage{listings}

\usepackage[unicode=true,bookmarks=true,
bookmarksnumbered=false,bookmarksopen=false,
breaklinks=false,
pdfborder={0 0 1},
backref=false,colorlinks=true]{hyperref}
\hypersetup{pdftitle={How to Recover a Qubit That Has Fallen Into a Black Hole}, pdfauthor={Aidan Chatwin-Davies, Adam S. Jermyn, and Sean M. Carroll}, citecolor=blue,linkcolor=blue,urlcolor=blue,citecolor=blue}

\usepackage[usenames,dvipsnames]{color}
\definecolor{purple}{rgb}{0.5,0,0.5}

\newcommand{\pr}{\prime}

\newcommand{\be}{\begin{equation}}
\newcommand{\ee}{\end{equation}}
\newcommand{\CG}[6]{\braket{{}^{#1}_{#2} \, {}^{#3}_{#4} | {}^{#5}_{#6}}}

\newcommand{\Eq}[1]{Eq.~\eqref{#1}}

\begin{document}

\preprint{\hbox{CALT-TH-2015-035}}

\title{How to Recover a Qubit That Has Fallen Into a Black Hole}
\date{\today}
\author{Aidan Chatwin-Davies}
\author{Adam S. Jermyn}
\author{Sean M. Carroll}
\thanks{\href{mailto:achatwin@caltech.edu}{\tt achatwin@caltech.edu}, \href{mailto:adamjermyn@gmail.com}{\tt adamjermyn@gmail.com},\\\href{mailto:seancarroll@gmail.com}{\tt seancarroll@gmail.com} \vspace{2mm}}
\affiliation{Walter Burke Institute for Theoretical Physics\\
California Institute of Technology,
Pasadena, CA 91125\\}

\begin{abstract}
We demonstrate an algorithm for the retrieval of a qubit, encoded in spin angular momentum, that has been dropped into a no-firewall black hole.
Retrieval is achieved analogously to quantum teleportation by collecting Hawking radiation and performing measurements on the black hole.
Importantly, these methods only require the ability to perform measurements from outside the event horizon.
\end{abstract}
\maketitle

\section{Introduction}

Recovering the complete quantum state of a black hole from the Hawking radiation \cite{Hawking1975} into which it evaporates is notoriously difficult \cite{Harlow2014}.
In this letter we tackle a simpler problem: recovering the quantum state of a single spin qubit that has fallen into an evaporating black hole.

\begin{figure}[ht]
\centering
\includegraphics[scale=0.75]{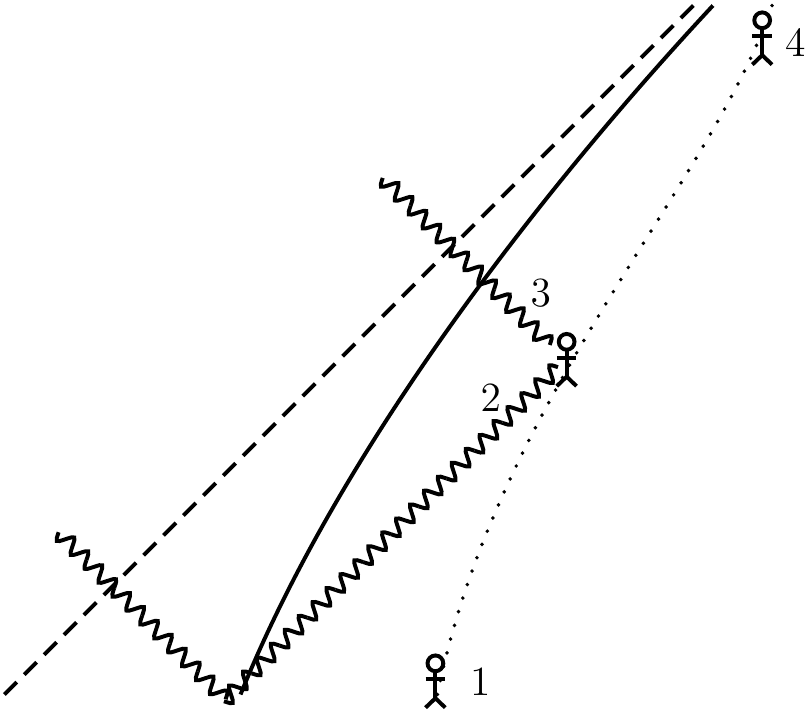}
\caption{Sketch of the qubit recovery protocol on a Penrose diagram. The numbers correspond to the steps enumerated above (details of the initial measurement 1. are not shown). The dashed line represents the event horizon, the solid line represents the stretched horizon \cite{Susskind1993}, and the dotted line represents the experimenter's trajectory.}
\label{fig:protocol}
\end{figure}

Our protocol uses information about the spin state of the black hole before and after the qubit entered, as well as the state of pairs of Hawking particles.
The outline of the procedure, sketched in Fig.~\ref{fig:protocol}, is as follows:
\begin{enumerate}
\item The initial spin state of the black hole is measured, putting the density matrix of the black hole in the form $\rho_B = \rho_B^{\mathrm{(int)}}\otimes |j,m\rangle\langle j,m|$, where $j,m$ are the quantum numbers for total and projected angular momentum, and $\rho_B^{\mathrm{(int)}}$ characterizes the internal degrees of freedom.
Perfect fidelity can be achieved only if $m=0$; the experimenter can measure the spin along different axes until this outcome is attained.
\item The experimenter collects a single Hawking photon that is part of a Bell pair, the other photon of which falls into the hole.
\item The qubit, a photon in an arbitrary helicity state $|\phi\rangle_A = \alpha\ket{\epsilon^+}_A + \beta \ket{\epsilon^-}_A$, is dropped into the hole.
\item The black hole's spin state is again measured, so that the density matrix becomes $\rho_B' = \rho_B^{\prime \mathrm{(int)}}\otimes |j',m'\rangle\langle j',m'|$. 
Dephasing of the hole's spin does not occur if the interactions between the hole's spin and its internal state are rotationally-invariant (conserve angular momentum).\footnote{Concretely, suppose that there was some conditional interaction between the black hole's internal degrees of freedom and its spin which would take a state $\ket{BH} \otimes (\alpha \ket{\epsilon^+} + \beta \ket{\epsilon^-})$ to a state $\alpha \ket{BH^+} \otimes \ket{\epsilon^+} + \beta \ket{BH^-}\otimes\ket{\epsilon^-}$, where $\braket{BH^+|BH^-} = 0$. If, for example, $\alpha = \beta = 1/\sqrt{2}$, then angular momentum in the $x$ direction would not be conserved by the interaction.}
\item The initial state of the qubit can then be reconstructed from the state of the collected Hawking photon.
\end{enumerate}
This falls far short of a resolution to the information-loss problem \cite{Page1993,Hawking2005,Susskind2005,Mathur2009}, but it does provide a concrete illustration of how information can escape from a black hole in certain special circumstances, and is similar in spirit to earlier discussions about using conserved quantities to recover black hole information \cite{Balasubramanian2006,Marolf2009}.
Moreover, whether or not the Page time \cite{Page2013} has elapsed does not affect information recovery, since the protocol is not concerned with reconstructing the state of the black hole.
In this regard the protocol is entirely distinct from the Hayden-Preskill result \cite{Hayden2007}.

\section{A protocol for retrieving individual qubits}

Suppose that Alice sits outside a black hole and has in her possession a photon in some state $\ket{\phi}_A = \alpha \ket{\epsilon^+}_A + \beta \ket{\epsilon^-}_A$ that is unknown to her.
Here, the basis states $\ket{\epsilon^+}_A$ and $\ket{\epsilon^-}_A$ represent the photon's helicity, and thus have angular momentum projection $+1$ and $-1$ respectively.
First, Alice measures the black hole's angular momentum and finds it in the state $\ket{j,m}_B$.
(We suppress the state of the black hole's internal degrees of freedom, which will play no role in our analysis.)
Such a measurement is technologically formidable, but one which Alice could in principle perform with the help of a sufficiently large Stern-Gerlach apparatus or by carefully measuring frame dragging.

Before dropping her qubit into the black hole, Alice collects a single Hawking photon.
We assume that the emitted photon is one half of a pair, the other one of which falls into the hole.
We also assume that Alice completes the protocol before any more Hawking particles are emitted.
The pairs of particles will have equal mass and opposite gauge and Poincar\'e quantum numbers.
Let us focus on angular momentum.

The states of photons with definite angular momentum are spherical waves that may be labelled by the quantum numbers for linear momentum, $k \in (0,\infty)$; total (spin plus orbital) angular momentum, $\eta \in \{1,2,\dots\}$; projected angular momentum, $\mu \in \{-\eta,...,\eta\}$; and parity, $\bar \omega \in \{+1,-1\}$ \cite{Messiah1965}.
We assume that the photons are each produced in the lowest angular momentum state ($\eta=1$) since this is the dominant mode of Hawking photon production.
Alternatively, Alice can measure her photon's total angular momentum and then discard her photon and restart the protocol if it does not have $\eta = 1$.
In order to preserve $CPT$, the two photons are produced with the same parity, since they are uncharged and since the wavefunctions of different parity for each $(k\eta\mu)$ have the same sign under $T$.
The photons must also be created in a zero total angular momentum state to conserve angular momentum.
As such, after Alice measures the parity of her photon, the angular momentum of the ingoing ($i$) and outgoing ($o$) Hawking photons is
\begin{align}
\nonumber \ket{0,0}_{io} &\equiv \frac{1}{\sqrt{3}}\left(\ket{1,1}_i \ket{1,-1}_o + \ket{1,-1}_i \ket{1,1}_o \right.\\
&\qquad\quad \left. - \ket{1,0}_i \ket{1,0}_o \right).
\end{align}
(Further justification for this model is provided in the next section.)

Next, Alice measures the squared projected angular momentum of her photon. If she obtains the result $\mu^2 = 0$, then she discards her photon and restarts the protocol.
Otherwise, the ingoing and outgoing photons are projected into the Bell state $\ket{\Phi}_{io} = (\ket{1,1}_i \ket{1,-1}_o + \ket{1,-1}_i \ket{1,1}_0)/\sqrt{2}$.
Finally, Alice drops in her qubit, and then measures the angular momentum of the hole again, determining it to be $\ket{j',m'}_B$.

After Alice collects a suitable Hawking photon and drops her qubit into the black hole, the total state of the black hole and the three photons is therefore $\ket{\Psi} = \ket{j,m}_B \otimes \ket{\phi}_A \otimes \ket{\Phi}_{io}$.
Alice is ignorant of what happens inside the black hole.
What Alice can know, however, is the total angular momentum of the black hole and the projection of its angular momentum vector along some axis.
As such, let us rewrite the $AiB$ subsystem in the total angular momentum basis:
\begin{widetext}
\begin{align} \label{eq:initialState}
\nonumber \ket{\Psi} &= \frac{1}{\sqrt{2}} \left\{ \sum_{\sigma=-2}^2 \left[ \CG{j}{m}{2}{2}{j+\sigma}{m+2} \ket{j+\sigma,m+2} \otimes \alpha \ket{1,-1}_o + \CG{j}{m}{2}{-2}{j+\sigma}{m-2} \ket{j+\sigma,m-2} \otimes \beta \ket{1,1}_o  \right] \right. \\
& \qquad + \frac{1}{\sqrt{6}} \sum_{\sigma=-2}^2 \left[ \CG{j}{m}{2}{0}{j+\sigma}{m} \ket{j+\sigma,m} \otimes (\alpha \ket{1,1}_o + \beta \ket{1,-1}_o) \right] \\
\nonumber & \qquad \left. + \frac{1}{\sqrt{2}} \sum_{\delta=-1}^1 \left[ \CG{j}{m}{1}{0}{j+\delta}{m} \ket{j+\delta,m}^\perp \otimes (\alpha \ket{1,1} - \beta \ket{1,-1}) \right] + \frac{1}{\sqrt{3}} \ket{j,m}^\vdash \otimes (\alpha \ket{1,1} + \beta \ket{1,-1}) \right\}
\end{align}
\end{widetext}
The symbols $\CG{j_1}{m_1}{j_2}{m_2}{j}{m} \equiv \braket{j_1,m_1;j_2,m_2|j,m}$ denote appropriate Clebsch-Gordan coefficients.
We have also suppressed the label $AiB$ on the total angular momentum kets.
Note that some of the $\CG{j_1}{m_1}{j_2}{m_2}{j}{m}$ could be zero.
For now, we will assume that $-j+2 < m < j-2$.
In particular note the following states: $\ket{j,m}$, which comes from $j \otimes 2$; $\ket{j,m}^\perp$, which comes from $j\otimes 1$; and $\ket{j,m}^\vdash$, which comes from $j \otimes 0$.
These states have the same angular quantum numbers, but are orthogonal.

Next, Alice queries the black hole's total angular momentum by performing the following orthogonal measurement on $AiB$:
\begin{equation}
\begin{array}{c}
\displaystyle \hat F_1 = \sum_a \ket{a,m}\bra{a,m}, \\
\displaystyle \hat F_2 = \sum_a \ket{a,m+2}\bra{a,m+2} + \ket{a,m-2}\bra{a,m-2}, \\
\displaystyle \hat F_3 = \hat I_{AiB} - \hat F_1 - \hat F_2.
\end{array}
\end{equation}
Note that by construction, only the results $\hat F_1$ and $\hat F_2$ may be obtained for black hole states which may emerge from this protocol.
The protocol for retrieving the state $\ket{\phi}$ is then as follows:

~

\emph{Case 1: Alice obtains the result $\hat F_1$.}
In this case, the whole system collapses to a state that is proportional to the second and third lines of \Eq{eq:initialState}.
Alice then measures the total angular momentum $\hat J^2$ of the black hole.

If Alice measures the result $J^2 = (j \pm 2)(j\pm 2 + 1)$, then she knows that the spin that she holds is in the desired state $\ket{\phi}_o = \alpha\ket{1,1}_o + \beta \ket{1,-1}_o$.

If Alice measures the result $J^2 = (j \pm 1)(j \pm 1 + 1)$, then the total system is in the state
\begin{align}
\nonumber \ket{\Psi^\pr} \propto & \frac{1}{\sqrt{6}} \CG{j}{m}{2}{0}{j\pm 1}{m} \ket{j \pm 1,m} \otimes \ket{\phi}_o \\
& \qquad + \frac{1}{\sqrt{2}} \CG{j}{m}{1}{0}{j\pm 1}{m} \ket{j \pm 1,m}^\perp \otimes \ket{\phi^\pr}_o,
\end{align}
while if she measures the result $J^2 = j(j+1)$, then the total system is in the state
\begin{align}
\nonumber \ket{\Psi^{\pr}} \propto & \left[ \frac{1}{\sqrt{6}} \CG{j}{m}{2}{0}{j}{m} \ket{j,m} + \frac{1}{\sqrt{3}} \ket{j,m}^\vdash\right] \otimes \ket{\phi}_o \\
& \qquad + \frac{1}{\sqrt{2}} \CG{j}{m}{1}{0}{j}{m} \ket{j,m}^\perp \otimes \ket{\phi^\pr}_o,
\end{align}
where $\ket{\phi^\pr}_o = \alpha\ket{1,1}_o - \beta \ket{1,-1}_o$.
Each of these states represents a mixed density matrix for the spin that Alice holds unless some of the Clebsh-Gordan coefficients vanish.
In particular, some algebra reveals that
\begin{equation} \label{eq:vanishing}
\begin{array}{lcl}
\CG{j}{m}{2}{0}{j+1}{m}^2 & = & \frac{3m^2(j+m+1)(j-m+1)}{j(j+1)(j+2)(2j+1)} \\[2mm]
\CG{j}{m}{2}{0}{j-1}{m}^2 & = & \frac{3m^2(j+m)(j-m)}{j(j+1)(j-1)(2j+1)} \\[2mm]
\CG{j}{m}{1}{0}{j}{m}^2 & = & \frac{m^2}{j(j+1)}.
\end{array}
\end{equation}
At the beginning of the protocol, Alice may measure $j$ and determine if it is an integer.
If not, she may repeatedly throw spin-$1/2$ particles into the black hole and measure $j$ until she measures an integral value.
She may then repeatedly measure the black hole's angular momentum projection along different axes until she obtains $m=0$, before collecting a Hawking photon and tossing her qubit into the hole.
In this way, the Clebsch-Gordan coefficients \eqref{eq:vanishing}  may be made to vanish, allowing Alice to recover the qubit.

~

\emph{Case 2: Alice obtains the result $\hat F_2$.}
In this case, the whole system collapses to a state that is proportional to the first line of \Eq{eq:initialState}.
Next, Alice measures the total angular momentum $\hat J^2$, obtaining the result $J^2 = (j+\sigma)(j+\sigma+1)$ for some $\sigma \in \{-2,\,\dots,2\}$.
The total state is then
\begin{align} \label{eq:case2}
\nonumber \ket{\Psi^{\pr\pr}} \propto ~ &\alpha \, \CG{j}{m}{2}{2}{j+\sigma}{m+2} \ket{j+\sigma,m+2} \otimes \ket{1,-1}_o \\
&- \beta \, \CG{j}{m}{2}{-2}{j+\sigma}{m-2} \ket{j+\sigma,m-2} \otimes \ket{1,1}_o .
\end{align}
We are faced with the problem of disentangling the $AiB$ part of the system from the $o$ part which Alice holds. She may accomplish this task with the help of a spin-2 ancilla and a local entangling unitary. Suppose Alice holds a spin-2 ancilla, $A^\pr$, that she prepares in the state $\ket{2,0}_{A^\pr}$.
If she then implements a local entangling unitary operator $U_{oA^\pr}$ such that
\begin{align} \label{eq:entanglingU}
\nonumber &U_{oA^\pr} \ket{1,1}_o \ket{2,0}_{A^\pr} = \ket{1,1}_o \ket{2,2}_{A^\pr} \\
&U_{oA^\pr} \ket{1,-1}_o \ket{2,0}_{A^\pr} = \ket{1,-1}_o \ket{2,-2}_{A^\pr},
\end{align}
upon acting with $U_{oA^\pr}$ on the spins that she holds, the total state $I_{AiB} \otimes U_{oA^\pr} \left(\ket{\Psi^{\pr\pr}} \otimes \ket{2,0}_{A^\pr}\right)$ is proportional to
\begin{align}
\nonumber &\alpha \, \CG{j}{m}{2}{2}{j+\sigma}{m+2} \ket{j+\sigma,m+2}_{AiB} \ket{1,-1}_o \ket{2,-2}_{A^\pr} \\
& - \beta \, \CG{j}{m}{2}{-2}{j+\sigma}{m-2} \ket{j+\sigma,m-2}_{AiB} \ket{1,1}_o \ket{2,2}_{A^\pr}.
\end{align}
Next, Alice tosses her ancilla into the black hole and then measures the black hole's total angular momentum.
The $AiBA^\pr$ terms will consist of linear combinations of $\ket{j+\sigma+2,m}$, $\,\dots$, $\ket{j+\sigma-2,m}$ weighted by the appropriate Clebsch-Gordan coefficients.
If Alice finds $AiBA^\pr$ in a total angular momentum $j+\sigma+\tau$ state, where $\tau \in \{-2,\,\dots,2\}$, it is straightforward to show that the spin that she still holds collapses to the state
\begin{align}
\nonumber \ket{\phi^{\pr\pr}}_o \propto~ &\alpha \, \CG{j}{m}{2}{2}{j+\sigma}{m+2}\CG{j+\sigma}{m+2}{2}{-2}{j+\sigma+\tau}{m} \ket{1,-1}_o \\
&- \beta \, \CG{j}{m}{2}{-2}{j+\sigma}{m-2}\CG{j+\sigma}{m-2}{2}{2}{j+\sigma+\tau}{m} \ket{1,1}_o .
\end{align}
As long as Alice measured the black hole angular momentum at the beginning of the protocol and ensured that $|m| \ll j$, then none of these coefficients vanish.
Alice then performs the appropriate unitary transformation on the spin that she holds to restore the state $\ket{\phi}_o$.

\section{Discussion}

We now consider several aspects of the proposed algorithm, as well as its consequences for black hole information theory.

~

\emph{State of the Hawking Photons:}
To see why the Hawking particles must be created in a zero total angular momentum state, note that spacetime is locally flat on the horizon and becomes increasingly flat as the black hole mass $M$ increases.
As a result, the only way for a Hawking pair to have non-zero angular momentum is for the pair to pick it up via interactions with the vacuum, \emph{i.e.}, with another Hawking pair.
This requires, roughly speaking, that two Hawking pairs be present within one wavelength $\lambda$ of one another in the time $t$ it takes for a pair to separate.
The relevant scaling relations in general are $\lambda  \propto T^{-1}$, $t \propto \lambda$, and $F \propto T^{d}$, where $d$ is the number of spatial dimensions, $T$ is the Hawking temperature, and $F$ is the particle number flux across the horizon.
The fraction $f$ of Hawking pairs which interact with an additional Hawking pair scales at tree order as $f \propto |\mathcal{A}|^2 (F \lambda^{d-1} t)^2 \propto |\mathcal{A}|^2$, where the mass-dependence of the phase-space factors dropped out.\footnote{This is not entirely unexpected. Consider, for instance, that the characteristic wavelength of Hawking photons is on the order of the Schwarzschild radius. Roughly speaking, since $t \propto \lambda$, any two photons at the black hole horizon will therefore overlap before they separate.}

For photons, which are the exponentially dominant form of Hawking radiation at large $M$, the matrix element $|\mathcal{A}|^2$ must depend on the probability of producing a virtual electron-positron pair to mediate the Hawking pair interaction.
This scales as $e^{-m_e/E_\gamma} \sim e^{-m_e GM}$.
Thus for large black holes, we expect these interactions to be exceedingly rare, and hence are justified in assuming that the photon pair carries no net angular momentum.
We note that the creation of Hawking pairs in the zero angular momentum state relies on the assumption that the local spacetime around the horizon of the black hole is a low-energy, quiescent environment.
Were there instead an energetic firewall at the horizon, we could not expect outgoing quanta to come from such a state.

When performing this analysis for other quantum numbers the same arguments apply: for large black holes, the Hawking pair must be created with zero net quantum number.
The algorithm we describe will work for any conserved quantum number which photons may carry, so long as the evolution of the relevant sector of the Hilbert space is unitary.
Notably, the algorithm does not require the hole's evolution in the total Hilbert space to be unitary over long timescales
If the relevant number is not quantized, the information recovered is only up to a precision limit given by the number of bits recovered.
For those quantum numbers which photons do not carry, superpositions of states cannot be recovered except by waiting exponentially long in $M$ for the relevant particles to be emitted.
If, on the other hand, it is known that a quantum number eigenstate fell in, and hence that only classical information was encoded in this way, then direct measurement of the black hole allows for recovery.
For example, in order to learn the mass of a particle that fell into the black hole, then one may of course measure the mass of the black hole afterwards, assuming that the initial mass of the black hole was known.
Altogether, this allows for unique recovery of classical information about any particle that fell in.
This is because each known fundamental particle has a unique set of gauge quantum numbers---mass, spin, charge, and color. 
This feature is not necessary---it would not hold in a theory with two unbroken $U(1)$ symmetries---but it does hold true in the Standard Model.

~

~

\emph{Resource Considerations:}
In its essence, our protocol amounts to a quantum teleportation scheme \cite{Bennett1993} between a transmitting party---the black hole---and a receiving party---Alice.
Its perfect fidelity when $m=0$ is due to the fact that setting $m=0$ eliminates any degeneracy in the states that the transmitting party could find after measuring in the total angular momentum basis, as opposed to a (nondegenerate) maximally-entangled basis. 
Alice would not be able to use an analogous procedure to recover more that a single qubit at a time, since the degeneracy of total angular momentum states rapidly increases as more and more spins are added.

We can also understand the difficulty of the multiple qubit case from the point of view of resources.
Suppose that Alice wishes to recover more than a single qubit at a time through a quantum number conservation protocol.
As these protocols amount to quantum teleportation schemes, Alice is bound by the resource inequality \cite{Wilde2013}
\begin{equation}
2[c\rightarrow c] + [qq] \geq [q\rightarrow q],
\end{equation}
which says that two classical bits, or cbits, of communication and one entangled qubit pair shared between the two parties is necessary to achieve one qubit of communication.
If Alice drops $N$ photons into the black hole and collects $N$ Hawking photons, she only obtains $\sim \log_2(N^2) = 2\log_2 N$ cbits since there are $4N+1$ possible outcomes for the total angular momentum measurement and $\sim 2N$ possible outcomes for the measurement of the projection of the angular momentum along the axis of quantization.
As such, she cannot hope to recover some general state of $N$ qubits, which would require $2N$ cbits.
On the other hand, she may be able to recover a state that is encoded in some subspace of $\mathcal{H}$.
For instance, Alice could try encoding her data in the total angular momentum of a set of $N$ qubits with total angular momentum $s$.
Thus she is encoding her data in a Hilbert space $\mathcal{H}_s$ with $\dim \mathcal{H}_s = 2s+1$.
Resource considerations do not prohibit the recovery of a state in $\mathcal{H}_s$, which only requires the extraction of $\log_2 \dim \mathcal{H}_s \leq \log_2(N+1)$ qubits and hence $\sim 2\log_2 N$ cbits.
We suspect that the general method for doing this is similar to the single qubit case.

~

\emph{Timescale Considerations:}
During the protocol, Alice must wait for the black hole to emit a quantum of Hawking radiation.
Hawking emission rates have been calculated by Page \cite{Page1976a}; for instance, photons are emitted in their lowest angular momentum mode at a rate given by $t_h^{-1} = 1.463 \times 10^{-4}~c^3/GM$ for Schwarzschild black holes.
Photon emission rates vary as a function of the black hole spin and can be on the order of one hundred times larger in the case of an extremal Kerr black hole \cite{Page1976b}, so let us express the timescale of Hawking emissions as $t_h = f \cdot GM/c^3$.
The factor $f$ contains both geometric and tunnelling factors, and is a function only of the spin of the black hole.

It is interesting to compare the emission time to the scrambling time \cite{Hayden2007,Shenker2014,Shenker2014a,Maldacena2015}, which may be thought of as the time it takes for Alice's infalling qubit to become incorporated into the (stretched horizon of) the black hole \cite{Susskind1993}.
The scrambling time is
\begin{equation}
t_s = \frac{1}{2\pi T} \ln S,
\end{equation}
where $S$ denotes the entropy of the black hole and where we have used units in which $\hbar$, $c$, and $k_B$ are 1.
This increases faster than $t_h$ as a function of the black hole radius $R$, since $S \propto R^2$ and $T \propto 1/R$, so there is a critical radius $R_\text{crit}$ above which the scrambling time is greater than the time required for a Hawking particle to be emitted.
In light of our single-qubit protocol, $R > R_\text{crit}$ means that the qubit which falls in is essentially bounced off of the black hole, rather than being incorporated into it.
The numerical factors involved, as well as the difference in scaling being in a logarithm, mean that the critical radius for a Schwarzschild black hole is very large ($R_\text{crit} \approx e^{853} l_p$, which is considerably larger than the current Hubble radius).
However, the dependence of $T$, $S$, and the numerical factors on spin means that this radius can be made arbitrarily small by tuning the angular momentum $J$ of the hole, since an extremal Kerr black hole has zero temperature but finite entropy.

\section{Conclusion}
We have described a protocol, based on quantum teleportation, that allows an external observer to recover a single spin qubit that has been dropped into a black hole, if the spin of the hole is measured before and after the qubit is dropped.
Our procedure relies on the fact that the angular momentum states of the black hole span the possible states of the qubit; for more than one qubit, this condition would not hold, and an analogous procedure would be unable to recover the information.
On the other hand, the fact that an external observer would see apparent information loss due to angular momentum state degeneracy is perhaps interesting in its own right.

This protocol retrieves a very specific kind of information: a single qubit encoded in a conserved quantity such as angular momentum; this is broad enough to include the information contained in any one particle within the Standard Model.
Importantly, it is the full quantum state of the qubit, not merely the classical angular momentum.
While our protocol does not extend to information encoded in the entanglement between multiple particles, the general idea of using quantum teleportation to recover information deserves further study.

~

\emph{Acknowledgements:} We would like to thank Gil Refael, who was instrumental in the conception of this protocol.
We also thank Ning Bao, Achim Kempf, Stefan Leichenauer, Don Marolf, Don Page, John Preskill, Grant Remmen, and Guillaume Verdon-Akzam for helpful discussions.
This research is funded in part by the Walter Burke Institute for Theoretical Physics at Caltech, by DOE grant DE-SC0011632, and by the Gordon and Betty Moore Foundation through Grant 776 to the Caltech Moore Center for Theoretical Cosmology and Physics.
ACD is supported by the NSERC Postgraduate Scholarship program. 
ASJ is supported by a Barry M. Goldwater Scholarship.

\bibliographystyle{utphys.bst}
\bibliography{refs}

\providecommand{\href}[2]{#2}\begingroup\raggedright\begin{thebibliography}{10}

\bibitem{Hawking1975}
S.~W. Hawking, ``{Particle creation by black holes},''
  \href{http://dx.doi.org/10.1007/BF02345020}{{\em Commun. Math. Phys.}
  {\bfseries 43} (1975) 199--220}.

\bibitem{Harlow2014}
D.~Harlow, ``{Jerusalem lectures on black holes and quantum information},''
  \href{http://arxiv.org/abs/1409.1231v2}{{\ttfamily arXiv:1409.1231v2}}.

\bibitem{Susskind1993}
L.~Susskind, L.~Thorlacius, and J.~Uglum, ``The stretched horizon and black
  hole complementarity,''
  \href{http://dx.doi.org/10.1103/PhysRevD.48.3743}{{\em Phys. Rev. D}
  {\bfseries 48} (1993) 3743--3761},
  \href{http://arxiv.org/abs/hep-th/9306069}{{\ttfamily arXiv:hep-th/9306069}}.

\bibitem{Page1993}
D.~N. Page, ``{Information in black hole radiation},''
  \href{http://dx.doi.org/10.1103/PhysRevLett.71.3743}{{\em Phys. Rev. Lett.}
  {\bfseries 71} no.~23, (1993) 3743},
  \href{http://arxiv.org/abs/hep-th/9306083}{{\ttfamily arXiv:hep-th/9306083}}.

\bibitem{Hawking2005}
S.~W. Hawking, ``Information loss in black holes,''
  \href{http://dx.doi.org/10.1103/PhysRevD.72.084013}{{\em Phys. Rev. D}
  {\bfseries 72} (2005) 084013},
  \href{http://arxiv.org/abs/hep-th/0507171}{{\ttfamily arXiv:hep-th/0507171}}.

\bibitem{Susskind2005}
L.~Susskind and J.~Lindesay, {\em An Introduction to Black Holes, Information
  and the String Theory Revolution: The Holographic Universe}.
\newblock World Scientific, 2005.

\bibitem{Mathur2009}
S.~D. Mathur, ``The information paradox: a pedagogical introduction,''
  \href{http://dx.doi.org/10.1088/0264-9381/26/22/224001}{{\em Class. Quantum
  Grav.} {\bfseries 26} no.~22, (2009) 224001},
  \href{http://arxiv.org/abs/0909.1038}{{\ttfamily arXiv:0909.1038}}.

\bibitem{Balasubramanian2006}
V.~Balasubramanian, D.~Marolf, and M.~Rozali, ``Information recovery from black
  holes,'' \href{http://dx.doi.org/10.1007/s10714-006-0344-8}{{\em Gen. Relat.
  Gravit.} {\bfseries 38} no.~11, (2006) 1529--1536},
  \href{http://arxiv.org/abs/hep-th/0604045}{{\ttfamily arXiv:hep-th/0604045}}.

\bibitem{Marolf2009}
D.~Marolf, ``Unitarity and holography in gravitational physics,''
  \href{http://dx.doi.org/10.1103/PhysRevD.79.044010}{{\em Phys. Rev. D}
  {\bfseries 79} (2009) 044010},
  \href{http://arxiv.org/abs/0808.2842}{{\ttfamily arXiv:0808.2842}}.

\bibitem{Page2013}
D.~N. Page, ``{Time dependence of Hawking radiation entropy},''
  \href{http://dx.doi.org/10.1088/1475-7516/2013/09/028}{{\em J. Cosmol.
  Astropart. Phys.} {\bfseries 2013} no.~09, (2013) 028},
  \href{http://arxiv.org/abs/1301.4995}{{\ttfamily arXiv:1301.4995}}.

\bibitem{Hayden2007}
P.~Hayden and J.~Preskill, ``Black holes as mirrors: quantum information in
  random subsystems,''
  \href{http://dx.doi.org/10.1088/1126-6708/2007/09/120}{{\em J. High Energy
  Phys.} {\bfseries 2007} no.~09, (2007) 120},
  \href{http://arxiv.org/abs/0708.4025}{{\ttfamily arXiv:0708.4025}}.

\bibitem{Messiah1965}
A.~Messiah, {\em Quantum Mechanics Volume II}.
\newblock North-Holland, 1965.

\bibitem{Bennett1993}
C.~H. Bennett, G.~Brassard, C.~Cr\'epeau, R.~Jozsa, A.~Peres, and W.~K.
  Wootters, ``{Teleporting an unknown quantum state via dual classical and
  Einstein-Podolsky-Rosen channels},''
  \href{http://dx.doi.org/10.1103/PhysRevLett.70.1895}{{\em Phys. Rev. Lett.}
  {\bfseries 70} (1993) 1895--1899}.

\bibitem{Wilde2013}
M.~Wilde, \href{http://dx.doi.org/10.1017/CBO9781139525343}{{\em Quantum
  Information Theory}}.
\newblock Cambridge University Press, 2013.

\bibitem{Page1976a}
D.~N. Page, ``{Particle emission rates from a black hole: Massless particles
  from an uncharged, nonrotating hole},''
  \href{http://dx.doi.org/10.1103/PhysRevD.13.198}{{\em Phys. Rev. D}
  {\bfseries 13} no.~2, (1976) 198--206}.

\bibitem{Page1976b}
D.~N. Page, ``{Particle emission rates from a black hole. II. Massless
  particles from a rotating hole},''
  \href{http://dx.doi.org/10.1103/PhysRevD.14.3260}{{\em Phys. Rev. D}
  {\bfseries 14} no.~12, (1976) 3260--3273}.

\bibitem{Shenker2014}
S.~H. Shenker and D.~Stanford, ``Black holes and the butterfly effect,''
  \href{http://dx.doi.org/10.1007/JHEP03(2014)067}{{\em J. High Energy Phys.}
  {\bfseries 2014} no.~3, (2014) 67},
  \href{http://arxiv.org/abs/1306.0622}{{\ttfamily arXiv:1306.0622}}.

\bibitem{Shenker2014a}
S.~H. Shenker and D.~Stanford, ``{Stringy effects in scrambling},''
  \href{http://arxiv.org/abs/1412.6087}{{\ttfamily arXiv:1412.6087}}.

\bibitem{Maldacena2015}
J.~Maldacena, S.~H. Shenker, and D.~Stanford, ``{A bound on chaos},''
  \href{http://arxiv.org/abs/1503.01409v1}{{\ttfamily arXiv:1503.01409v1}}.

\end{thebibliography}\endgroup

\end{document}